\documentclass{article}

\usepackage{arxiv}
\usepackage[utf8]{inputenc} 
\usepackage[T1]{fontenc}    
\usepackage{hyperref}       
\usepackage{url}            
\usepackage{booktabs}       
\usepackage{amsfonts}       
\usepackage{nicefrac}       
\usepackage{microtype}      
\usepackage{lipsum}		
\usepackage{graphicx}
\usepackage{natbib}
\usepackage{doi}
\usepackage{multirow}
\usepackage{authblk}
\usepackage[symbol]{footmisc}
\usepackage{lineno}
\usepackage{tabularx}
\usepackage{xcolor}
\usepackage{marvosym}
\usepackage{ifsym}

\setcitestyle{numbers,square}

\title{A foundation model for generalizable disease diagnosis in chest X-ray images}

\author[\Letter,1,2]{Lijian~Xu}
\author[3]{Ziyu~Ni}
\author[1]{Hao~Sun}
\author[1,4]{Hongsheng~Li} 
\author[\Letter,2]{Shaoting~Zhang}

\affil[1]{Centre for Perceptual and Interactive Intelligence, the Chinese University of Hong Kong, Hong Kong}
\affil[2]{Shanghai Artificial Intelligence Laboratory, Shanghai}
\affil[3]{Sensetime Research, Shanghai}
\affil[4]{Department of Electronic Engineering, the Chinese University of Hong Kong, Hong Kong}

\date{}

\begin{document}
\maketitle

\begin{abstract}
Medical artificial intelligence (AI) is revolutionizing the interpretation of chest X-ray (CXR) images by providing robust tools for disease diagnosis. However, the effectiveness of these AI models is often limited by their reliance on large amounts of task-specific labeled data and their inability to generalize across diverse clinical settings. To address these challenges, we introduce CXRBase, a foundational model designed to learn versatile representations from unlabelled CXR images, facilitating efficient adaptation to various clinical tasks.
CXRBase is initially trained on a substantial dataset of 1.04 million unlabelled CXR images using self-supervised learning methods. This approach allows the model to discern meaningful patterns without the need for explicit labels. After this initial phase, CXRBase is fine-tuned with labeled data to enhance its performance in disease detection, enabling accurate classification of chest diseases.
CXRBase provides a generalizable solution to improve model performance and alleviate the annotation workload of experts to enable broad clinical AI applications from chest imaging.

\end{abstract}


\section{Introduction}
Chest X-ray is a commonly used imaging technique in everyday clinical practice, providing valuable insights into thoracic diseases. In many cases, additional examinations are necessary to establish a differential diagnosis. However, in emergency settings or facilities with high patient volumes, there is a need for a rapid screening and reporting system to improve the efficiency of the clinical workflow.
Although there are several advanced AI models available for chest X-ray diagnosis \cite{xu2024medvilam}, the development of a foundational model that can be easily adapted to various downstream tasks remains a critical need.

Image-based self-supervised training methods have been popular recently and can be categorized into several approaches: innate relationships \cite{Rotations2018unsupervised}, generative methods \cite{2014generative}, contrastive techniques \cite{SimCLR2020Contrastive}, and self-prediction methods \cite{huang2023self}.
Among these, the introduction of Vision Transformers (ViT) has led to the development of Masked Autoencoders (MAE) \cite{he2022masked}, which leverage self-prediction techniques and have demonstrated exceptional performance across various natural image tasks.
However, in the realm of medical imaging, self-masking approaches pose challenges by potentially altering crucial semantic information within images \cite{huang2023self}. This issue arises because medical images often display macroscopic similarities, while significant pathological features are localized, making them more sensitive to changes introduced by self-masking.
In general images, the AttMask strategy \cite{eccv2022attmask} utilizes attention maps to inform the masking process based on significant local features. However, applying this approach to medical images for the Masked Autoencoder (MAE) task presents challenges in directly extracting these important local features.
To address this, incorporating medical domain knowledge, such as anatomical information, into a self-supervised framework can improve the transferability of downstream tasks, particularly when labeled data is scarce \cite{zhou2023advancing, wu2023medklip, CheXzero2022}.
In this context, the report-supervised R2L method was introduced \cite{zhang2022contrastive} to obtain supervision from radiology reports. This approach leverages the words and sentences in free-text reports as guidance for deep neural networks, enabling them to learn effective radiograph representations. It has been shown to outperform traditional label-based and self-supervised pre-training methods by significant margins across various downstream tasks \cite{zhou2021models, zhang2022contrastive}.

In this study, we introduced a novel SSL-based foundation model for chest X-ray (CXR) images, called CXRBase, and systematically evaluated its performance and generalizability across disease screening and diagnosis tasks. 
We constructed CXRBase from large-scale unlabelled CXR images using SSL methods (i.e., masked autoencoder), applied consecutively on both natural images (ImageNet-1k) and CXR images from public datasets, including a total of 1.04 million CXR images. 
Additionally, we supplemented the training with our in-house dataset.
We adapted CXRBase to a series of challenging screening and disease diagnosis tasks by fine-tuning CXRBase with specific task labels and then validate its performance. 
CXRBase achieves consistently superior performance and label efficiency in adapting to these tasks, compared to competing specialist models, including that pre-trained on ImageNet-21k with traditional transfer learning. 
We further explored the interpretability of CXRBase's disease localization performance through qualitative results and controlled experiments, highlighting how salient image regions align with established knowledge from the chest AI literature.

\begin{figure}[t]
	\centering
        \includegraphics[width=0.95\linewidth]{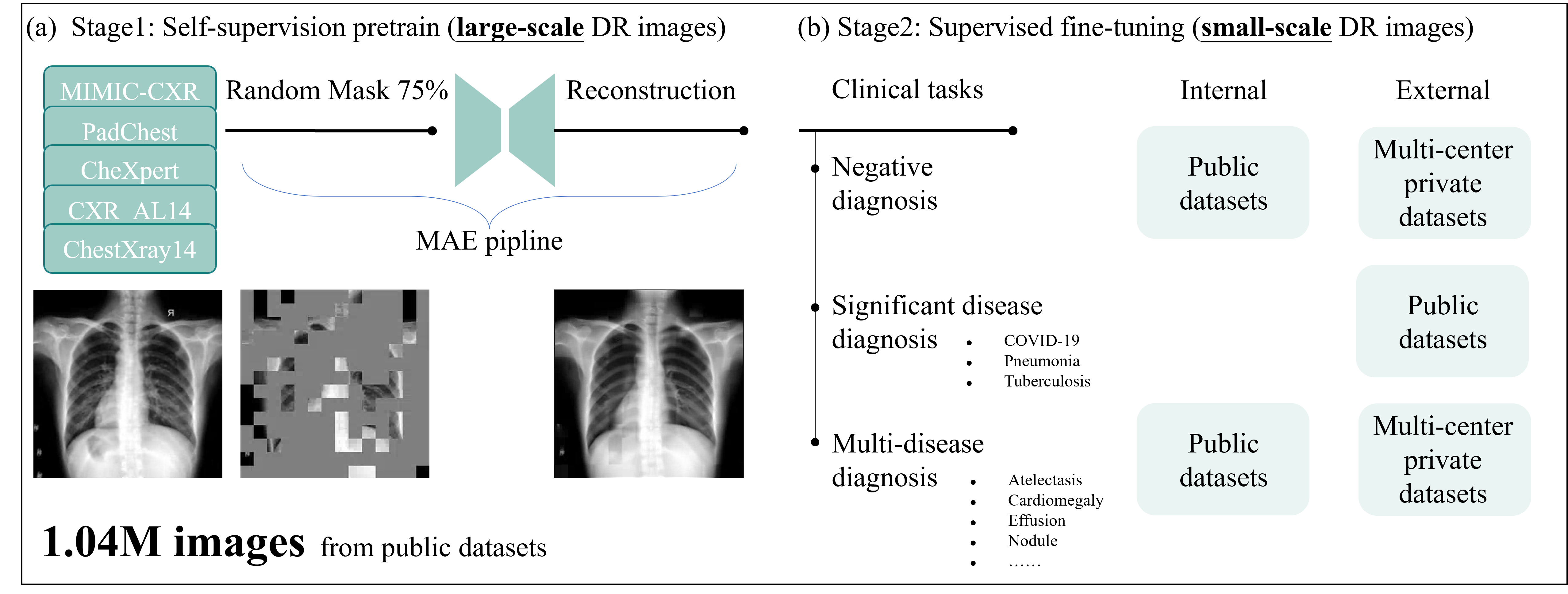}
	\caption{Schematic of our proposed foundation model for Chest X-ray .}
	\label{fig:overview}
\end{figure}

\begin{figure}[t]
	\centering
        \includegraphics[width=0.95\linewidth]{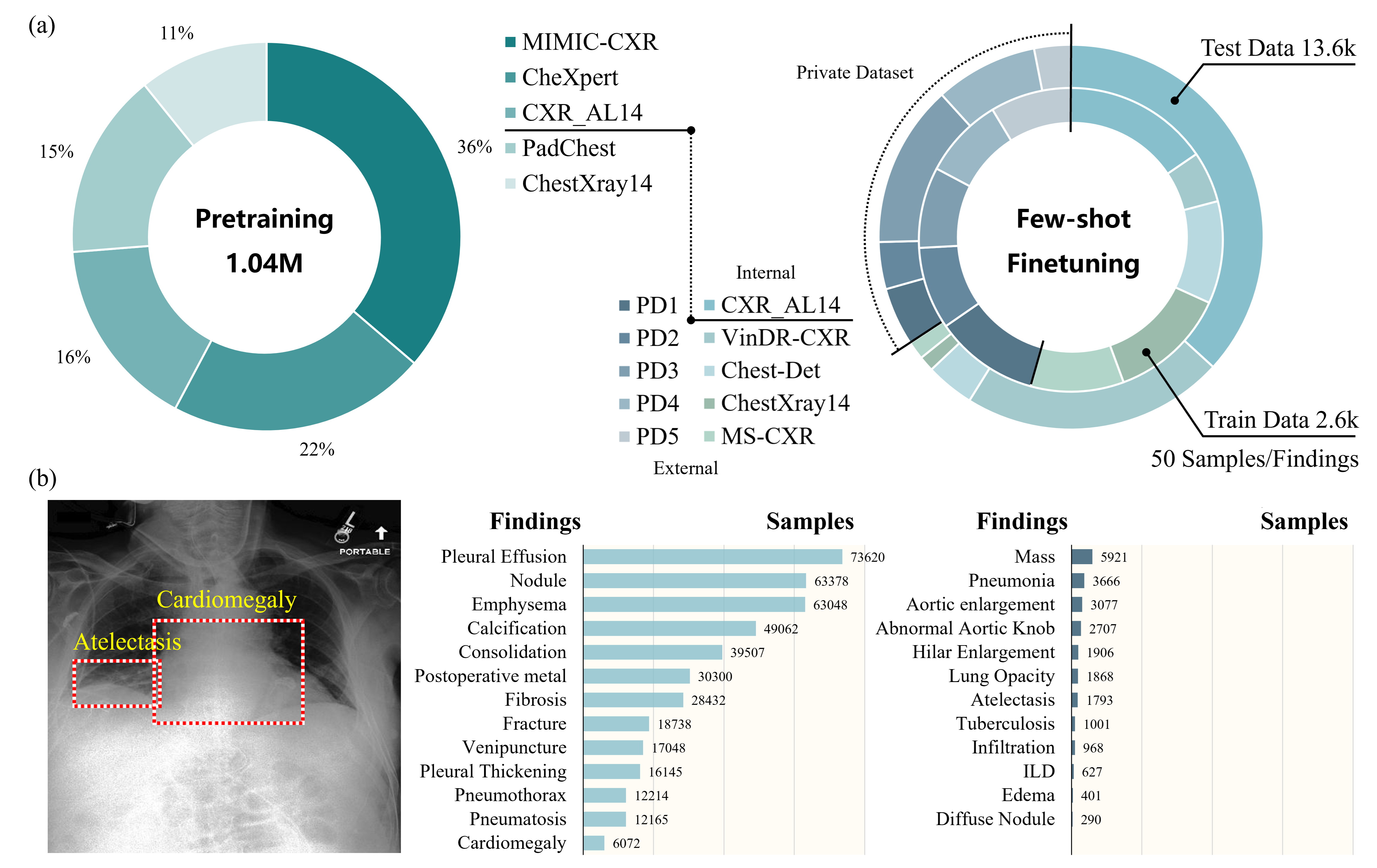}
	\caption{(a) Dataset distribution of pretraining and downstream task finetuning. (b) Data samples of Disease Localization task on 25 classes (total 453,954 samples). It can be observed that the data tend to show a long-tailed distribution. }
	\label{fig:dataset}
\end{figure}

\section{Results}

\subsection{Overview}

Figure \ref{fig:overview} provides an overview of the construction and application of CXRBase. 
We developed a foundation model by self-supervision pretraining on a large-scale DR images.
To build CXRBase, we assembled a dataset of 1.04 million CXR images, with 30.5\% of images sourced from private datasets and 69.5\% from public datasets (see Figure \ref{fig:dataset}). 
After pretraining CXRBase using self-supervised on these CXR images, we extensively evaluated its performance and generalizability in adapting to various chest disease tasks. 
For the tasks of chest disease screening and diagnosis, we selected multi-center private datasets and publicly available datasets.

\begin{figure}[t]
	\centering
        \includegraphics[width=0.95\linewidth]{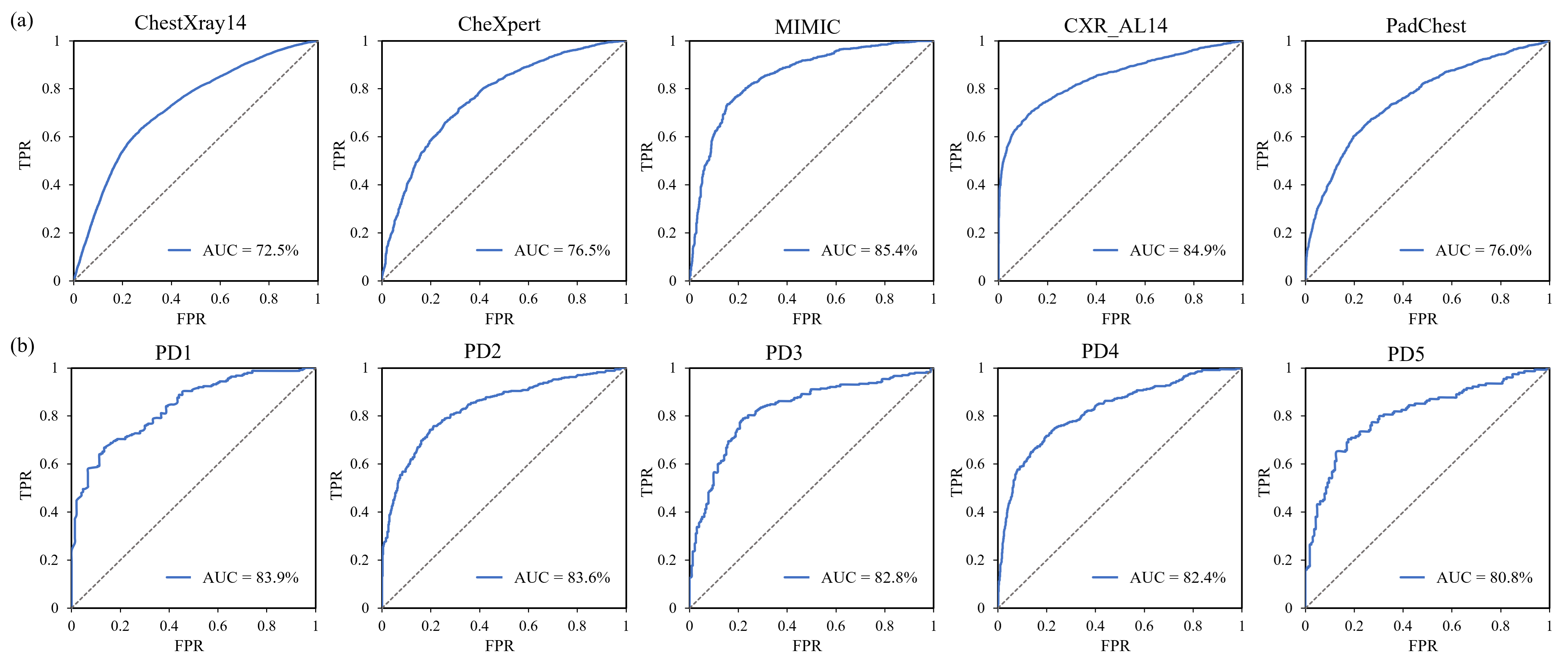}
	\caption{Model performance of \textbf{Negative Screening task} on 10 test sets from five public large datasets (a) and five private datasets (b) of Chest X-ray images. The public datasets comprise MIMIC, CheXpert, PadChest, ChestXray14 and CXR-AL14, while the private datasets include the retrospective and prospective images from three centers and two centers, respectively.}
	\label{fig:neg_screening}
\end{figure}



\subsection{Screening}

\textbf{Negative Screening} We conducted negative screening, i.e., binary classification on five private and five public datasets for chest X-ray. 
The public datasets comprise MIMIC, CheXpert, PadChest, ChestXray14
and CXR-AL14, while the private datasets include the retrospective and prospective images from three centers and two
centers, respectively.
Figure \ref{fig:neg_screening} shows that the proposed model achieves an averaged AUC of 79.1\%  and 82.7\% for the five public datasets and the five private datasets, respectively.
Detailed comparative results with could be available in the supplement material.

\begin{figure}[t]
	\centering
        \includegraphics[width=0.95\linewidth]{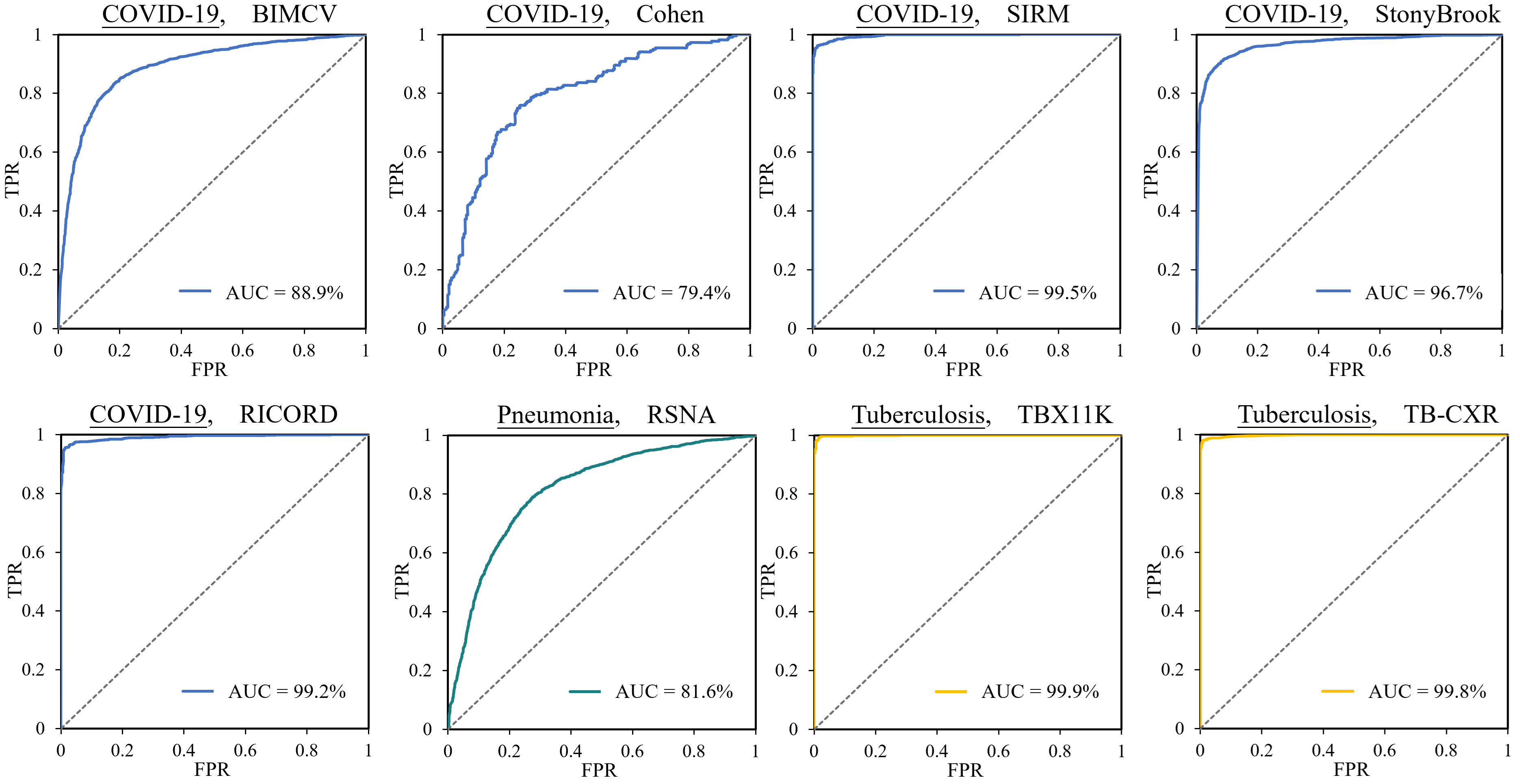}
	\caption{
 Model performance of \textbf{Screening task} on eight single-class disease datasets. The datasets include three important disease categories: COVID-19, pneumonia, and tuberculosis. 
 ACC and AUC are used as evaluation metrics. 
}
	\label{fig:cls_single}
\end{figure}

\textbf{COVID/Tuberculosis Screening} We employed eight publicly available datasets to verify the performance of CXRBase on various chest diseases, as depicted in Figure \ref{fig:cls_single}. 
In COVID-19 classification, CXRBase achieved high AUROC values of 88.9\% (95\% CI 83.1-99.4\%), 79.4\% (95\% CI 72.9-96.7\%), 99.5\% (95\% CI 92.9-96.7\%), 96.7\% (95\% CI 92.9-96.7\%)  and 99.2\% (95\% CI 88.0-99.7\%) on BIMCV, Cohen, SIRM, StonyBrook and RICORD datasets, respectively. 
Similar superior performance was observed for the pneumonia and tuberculosis classification. 
The AUPR results of CXRBase were also significantly higher than the compared groups. 
For instance, when fine-tuned on Cohen, CXRBase achieved AUROC of 96.2\% (95\% CI 91.5-98.9\%) and 
93.8\% (95\% CI 92.9-94.7\%), respectively, on SIRM and BIMCV datasets, statistically significantly higher than SL-ImageNet
 (\textit{p} < 0.001) on SIRM and SSL-ImageNet (\textit{p} < 0.001) on BIMCV. 
 For detailed quantitative results, please refer to Supplementary Table 1.

\begin{table}[]
\caption{ Model performance of \textbf{Disease Classification} task with few-shot inference setting on five public datasets and five private datasets. The mean metrics (e.g., AUC and F1) refer to the macro average on each disease. }
    \centering
    \begin{tabularx}{\textwidth}{cc*{15}{X}}
\hline
 Dataset & Metric & \rotatebox{90}{Mean}& \rotatebox{90}{Atelectasis} & \rotatebox{90}{Cardiomegaly} & \rotatebox{90}{Effusion} & \rotatebox{90}{Nodule} & \rotatebox{90}{Pneumonia} & \rotatebox{90}{Pneumothorax}  &\rotatebox{90}{Consolidation} &\rotatebox{90}{Edema} &\rotatebox{90}{Emphysema} &\rotatebox{90}{Fibrosis} & \rotatebox{90}{Pleural Thicken} & \rotatebox{90}{Fracture} & \rotatebox{90}{Tuberculosis} & \rotatebox{90}{Hilar Enlargement} \\
\midrule
MIMIC& AUC &74.7	&74.5	&74.5	&88.1	&/	&62.3	&78.5	&74.4	&83.2	&/	&/	&/	&61.9	&/	&/\\
& F1&38.3	&52.1	&49.3	&68.3	&/	&30.9	&24.1	&22.5	&52.6	&/	&/	&/	&6.7	&/	&/\\
CheXpert& AUC &71.7	&65.1	&79.6	&81.9	&/	&64.7	&79.7	&60.6	&79.7	&/	&/	&/	&62.5	&/	&/\\
& F1  &43.8	&51.3	&47.2	&72.9	&/	&27.2	&41.9	&35.7	&59.7	&/	&/	&/	&14.3	&/	&/ \\
ChestXray14& AUC &75.4	&75.1	&80.6	&86.0	&66.1	&68.1	&78.9	&77.2	&83.3	&76.1	&69.5	&68.4	&/	&/	&/\\
& F1  &20.8	&30.5	&21.7	&48.6	&17.0	&5.4	&30.9	&20.1	&15.2	&16.6	&7.6	&14.7	&/	&/	&/ \\
CXR-AL14& AUC &80.5	&95.9	&/	&93.7	&58.2	&/	&93.4	&85.2	&/	&93.1	&69.3	&78.8	&56.9	&/	&/\\
& F1  &48.9	&47.8	&/	&83.2	&35.5	&/	&60.1	&66.9	&/	&70.0	&30.4	&34.6	&12.0	&/	&/\\
PadChest& AUC &82.5	&71.7	&88.2	&93.5	&67.1	&79.9	&91.9	&85.4	&95.6	&81.6	&93.7	&76.9	&69.2	&93.0	&66.6\\
& F1  &26.4	&18.3	&50.8	&63.8	&12.3	&31.7	&17.6	&18.5	&22.3	&14.8	&34.9	&25.5	&11.0	&33.3	&15.1\\
PD1& AUC &79.9	&87.0	&91.4	&91.4	&64.2	&82.6	&87.4	&84.4	&97.4	&82.4	&55.4	&66.1	&66.5	&88.6	&74.1\\
& F1  &48.1	&52.6	&72.7	&81.1	&37.8	&59.6	&53.8	&15.4	&64.5	&37.6	&30.6	&31.7	&32.1	&54.8	&48.8\\
PD2& AUC &83.3	&91.7	&90.7	&92.9	&66.9	&83.1	&90.8	&87.9	&97.3	&87.0	&69.4	&72.9	&75.6	&83.6	&76.5\\
& F1  &40.4	&38.2	&67.3	&71.7	&39.7	&56.3	&44.4	&10.0	&48.2	&31.7	&29.0	&27.6	&25.1	&40.5	&35.4\\
PD3& AUC &81.5	&89.5	&88.0	&88.5	&64.5	&82.8	&80.3	&92.2	&96.4	&87.7	&72.9	&71.9	&69.2	&85.0	&72.1\\
& F1  &33.1	&14.0	&68.5	&61.6	&23.7	&52.3	&16.2	&7.5	&37.2	&32.0	&23.5	&24.8	&24.8	&42.3	&34.3\\
PD4& AUC &90.2	&94.7	&95.1	&92.9	&62.1	&86.0	&98.4	&96.7	&99.2	&97.5	&86.6	&92.4	&90.2	&76.0	&95.5\\
& F1  &27.7	&15.4	&64.7	&41.0	&20.3	&48.0	&13.3	&7.1	&66.7	&20.7	&17.2	&18.2	&16.7	&7.8	&30.8\\
PD5& AUC &86.6	&99.7	&92.2	&99.3	&55.2	&82.9	&/	&/	&70.1	&97.5	&86.1	&91.8	&/	&85.2	&93.0\\
& F1  &40.3	&66.7	&57.7	&66.7	&27.6	&48.8	&/	&/	&3.4	&44.4	&19.6	&28.6	&/	&71.4	&8.7\\
		\bottomrule
    \hline
\end{tabularx}
    \label{tab:cls}
\end{table}


\subsection{Disease Diagnosis}
\textbf{Multi-label Classification} We employed ten publicly available datasets to verify the performance of CXRBase on 14 chest diseases, 
as depicted in Table \ref{tab:cls}. 
CXRBase consistently demonstrated the best performance across most datasets compared to other specialist models. 
For the public datasets, CXRBase achieves an averaged AUC and F1 of 76.9\% and 35.6\%,respectively.
We further evaluated the generalizability on the external datasets and found that the proposed model
attained an averaged AUC and F1 of 84.3\% and 37.9\% on the five private datasets. 
For detailed quantitative results, please refer to Supplementary Table 2.

\textbf{Disease Localization}
We evaluated the performance of localization on 15 chest diseases using five private and five public datasets. CXRBase achieved an average AP50 of 19.2\% on the private datasets and 19.1\% on the public datasets. For detailed quantitative results comparing CXRBase with other advanced specialist models, please refer to Supplementary Table 3.


\begin{table}[]
\caption{Model performance of \textbf{Disease Localization} task with 50-shot setting on five public datasets (MS-CXR, VinDr-CXR, ChestXray14, CXR-AL14, Chest-Det) and five private datasets. The metric (AP50)  refers to the macro average on the 15 diseases.}
    \centering
    \begin{tabularx}{\textwidth}{*{2}{c}*{15}{X}}
\hline
  Dataset & Mean & \rotatebox{90}{Atelectasis} & \rotatebox{90}{Cardiomegaly} & \rotatebox{90}{Effusion}  &\rotatebox{90}{Nodule}&\rotatebox{90}{Pneumonia} &\rotatebox{90}{Pneumothorax} &\rotatebox{90}{Consolidation} &\rotatebox{90}{Edema} &\rotatebox{90}{Emphysema} &\rotatebox{90}{Fibrosis} &\rotatebox{90}{Pleural Thickening} &\rotatebox{90}{Fracture} &\rotatebox{90}{Tuberculosis} &\rotatebox{90}{Hilar Enlargement} &\rotatebox{90}{Abnormal Aortic Knob} \\
\midrule
MS-CXR & 19.9 & 14.0 & 66.6 & 12.8 & / & 14.8 & 13.7 & 10.0 & 7.1 & / & / & / & / & / & / & /
\\
VinDr-CXR & 15.5 & 9.0 & 44.1 & 13.5 & 6.9 & / & 10.4 & 6.8 & / & / & 7.4 & 6.7 & / & / & / & 34.5
\\
ChestXray14 & 15.9 & 6.5 & 60.3 & 5.6 & 8.2 & 6.9 & 7.8 & / & / & / & / & / & / & / & / & /
\\
CXR-AL14 & 24.5 & 22.3 & / & 20.1 & 10.6 & / & 34.1 & 24.1 & / & 54.5 & 16.3 & 11.7 & 26.6 & / & / & /
\\
Chest-Det & 19.8 & 12.1 & 56.2 & 18.3 & 6.0 & / & 17.2 & 26.4 & / & 29.7 & 14.9 & 11.2 & 5.7 & / & / & /
\\
PD1 & 27.6 & 29.7 & 70.3 & 32.5 & 11.6 & 17.7 & 31.2 & 5.0 & 52.0 & 10.6 & 15.9 & 6.7 & 15.5 & 40.1 & 32.7 & 43.1
\\
PD2 & 22.9 & 16.0 & 59.2 & 20.9 & 9.6 & 11.0 & 26.5 & 6.8 & 44.1 & 15.8 & 10.2 & 5.7 & 12.9 & 42.4 & 19.2 & 43.1
\\
PD3 & 17.7 & 5.0 & 37.5 & 23.7 & 18.9 & 12.0 & 5.0 & / & 7.6 & 10.6 & 8.5 & 5.7 & 5.8 & 44.3 & 8.6 & 54.2
\\
PD4 & 16.1 & 5.1 & 20.6 & 13.3 & 6.1 & 6.0 & 105.0 & / & 5.0 & 5.2 & 5.1 & 5.6 & 5.8 & 13.5 & 20.4 & 8.8
\\
PD5 & 11.7 & 5.2 & 26.3 & 8.7 & 8.9 & 7.2 & / & / & 5.0 & 5.1 & 4.9 & 5.0 & / & 24.8 & 5.8 & 33.1
\\
		\bottomrule
\hline
\end{tabularx}
    \label{tab:VG}
\end{table}








\section{Material and Method}
\subsection{Datasets for Pre-training}

We trained the model on multiple public datasets.

\begin{itemize}
\item \textbf{MIMIC-CXR} \cite{johnson2019mimic} contains more than 377,110 radiograph images from over 227,835 radiographic studies. 
Each radiograph is paired with lesion classification information and associated radiology report. We employ this dataset for multi-label classification and report generation tasks.
\item \textbf{Padchest} \cite{padchest} includes 160,840 images obtained from 67,000 patients, covering six different position views with related reports. The reports were labeled with 174 different radiographic findings which can be used for classification task.
\item \textbf{ChestXray14} \cite{wang2017chestx} is an available dataset for diagnosing 14 common lung diseases and eight localization of key findings, with 984 radiograph images and hand-labeled BBOX. We randomly split it into training/validation/test sets by 7:1:2 for classification and localization task.
\item \textbf{CheXpert} \cite{irvin2019chexpert}contains 224,316 images collected from 65,240 patients. We extract 1\% of the dataset to conduct a finetuning experiment for multi-diseases classification. We follow MRM to focus on 5 diseases: Atelectasis, Cardiomegaly, Consolidation, Edema and Pleural Effusion. We sample training/test sets from the official training set and they constitutes 21,84/5,000 images of the whole dataset. For the CheXpert dataset, we randomly sample 5,000 images with the same 14 labels with the MIMIC-CXR dataset for testing. 


\end{itemize}

\subsection{Downstream finetuning and evaluation}


\textbf{Ethical statement}
Our dataset consists of original images derived from public available datasets. It adheres to stringent ethical guidelines, as detailed in each dataset. 
We also evaluate the model performance on five private datasets from different centers, where three datasets were retrospectively derived and the two datasets were prospectively collected. The retrospective study was approved by the Ethics Committee of Fengcheng People's Hospital, Huanggang Hospital, and Longkou People's Hospital.
and the committee waived the consent since the retrospective research would not change the examination process of the patients. All data were adequately anonymized. 

\begin{itemize}
\item \textbf{VinDr-CXR} \cite{nguyen2022vindr} includes chest radiographs with annotations for the classification of 28 common chest diseases. The dataset contains 15,000 CXR scans in the training set. Each scan is annotated by three radiologists. We select eight diseases from the dataset along with their corresponding BBox for the disease localization task.
\item \textbf{ChestX-Det} \cite{lian2021structure} consists of 3,578 images from NIH ChestXray14\cite{wang2017chestx}, which are annotated by three radiologists for 13 common disease or abnormality categories. We select seven diseases from the dataset along with BBox for the disease localization task.
\end{itemize}

\textbf{Pneumonia and COVID-19 Classification}
COVIDx CXR-4\cite{wang2020covid} is an open-source and continuously updated dataset of COVID-19 chest images, which contains 84,818 images from 45,342 subjects. We conducted separate tests on 6 image sources included in this dataset, which consist of COVID-19 or common pneumonia images, to cross-validate the performance of the model. The description of the relevant dataset is as follows:
\begin{itemize}
\item \textbf{covid-chestxray-dataset} \cite{cohen2020covid} (referred to as ‘Cohen’ based on the author's name for the sake of simplicity) is a public open dataset of chest X-ray and CT images of patients which are positive or suspected of COVID-19 or other viral and bacterial pneumonia (MERS, SARS, and ARDS). 
\item \textbf{COVID-19 Radiography Database} \cite{chowdhury2020can,rahman2021exploring} is a database of COVID-19 x-ray images from the Italian Society of Medical and Interventional Radiology (SIRM) COVID-19 DATABASE.
\item \textbf{BIMCV-COVID19+} \cite{vayá2020bimcv} is a large dataset with chest X-ray images CXR (CR, DX) and computed tomography (CT) imaging of COVID-19 patients along with their radiographic findings, pathologies, polymerase chain reaction (PCR), immunoglobulin G (IgG) and immunoglobulin M (IgM) diagnostic antibody tests and radiographic reports from Medical Imaging Databank in Valencian Region Medical Image Bank (BIMCV).
\item \textbf{saltz2021stony} \cite{saltz2021stony} collect the cases at Stony Brook University from patients who tested positive for COVID-19. The collection includes images from different modalities and organ sites (chest radiographs, chest CTs, brain MRIs, etc.).
\item \textbf{RSNA International COVID-19 Open Radiology Database} \cite{tsai2021rsna} (RICORD) was created through a collaboration between the RSNA and the Society of Thoracic Radiology (STR). Clinical annotation by thoracic radiology subspecialists was performed for all COVID-positive chest radiography (CXR) imaging studies.
\item \textbf{RSNA Pneumonia} \cite{shih2019augmenting} is a binary classification chest X-ray dataset, where each radiograph is categorized as either pneumonia or normal. We randomly sample 3,000 data from the official
training set to build the test set for direct inference and finetuning experiments for disease classification and localization tasks.
\end{itemize}

\textbf{Tuberculosis Classification}
\begin{itemize}
\item \textbf{TBX11K} \cite{liu2020tbx11k} is a large dataset containing 11000 chest x-ray images. It's the only TB dataset that includes TB bounding boxes. This allows both classification and localization models to be trained.
\item \textbf{TB Chest X-ray} \cite{tb9224622} is a database of chest X-ray images for Tuberculosis (TB) positive cases along with Normal images. There are 700 TB images and 3500 normal images, publicly accessible in kaggle.
\end{itemize}

\subsection{Model Architecture}

\textbf{Data processing}
We excluded the background and retain only the chest area of the images. All the images are resized to a resolution of 512 × 512 using cubic interpolation.
During the model training process, we apply the same data augmentation techniques as the masked autoencoder. These techniques include random cropping, where the lower bounds for cropping are set to 20\% of the whole image, and the upper bounds are set to 100\%. The cropped patches are resized to 224×224. Additionally, random horizontal flipping is applied to the images, and image normalization is performed as part of the augmentation process.

\textbf{Implementation}
We employed a specific configuration of the masked autoencoder, which comprises an encoder and a decoder. 
The encoder component utilizes a large vision Transformer (ViT-large) with 24 Transformer blocks. The embedding vector size is set to 1,024. On the other hand, the decoder component consists of a small vision Transformer (ViT-small) with eight Transformer blocks. The embedding vector size for the decoder is set to 512.
During the training process, the encoder takes unmasked patches of size 16×16 as input and projects them into a feature vector of size 1,024. The 24 Transformer blocks, which include multiheaded self-attention and multilayer perceptron layers, process the feature vectors to generate high-level features.
The decoder inserts masked dummy patches into the extracted high-level features as input to the model. It then aims to reconstruct the image patch after a linear projection.
In model training, the objective is to reconstruct CXR images from highly masked versions. The mask ratio is set to 0.75 for CXR images. The batch size is set to 2,048, which is distributed across eight GPUs with 224 samples per GPU.
The total training epoch is set to 800, with the first 30 epochs dedicated to learning rate warming up. 
At the end of the training process, the model weights from the optimal epoch are saved as the checkpoint for adapting to downstream tasks.

\textbf{Adaptation to downstream tasks}
For downstream tasks, we solely utilized the encoder (ViT-large) from the foundation model and discard the decoder. This encoder generates high-level features from CXR images. These features are then fed into a multilayer perceptron, which produces probabilities for different disease categories. The final classification is determined by selecting the category with the highest probability. The number of categories determines the number of neurons in the final layer of the multilayer perceptron.
To prevent overfitting and regulate the output distribution, we incorporate label smoothing by softening the ground-truth labels in the training data. 
The goal of training is to generate categorical outputs that match the provided labels. We use a batch size of 16 during training and conduct a total of 50 epochs. 
After each epoch of training, the model is evaluated on the validation set. The model checkpoint with the highest AUROC on the validation set is saved. This checkpoint serves for both internal and external evaluation of the model's performance.

\section{Related Work}
\subsection{Self-supervised Represent Learning}
\textbf{Contrastive}
The self-supervised methods are built on the principle that variations induced by transforming an image do not alter its underlying semantic meaning. Therefore, different augmentations of the same image are considered positive pairs, while other images and their augmentations are regarded as negative pairs with respect to the current instance. The objective is to train a model that minimizes the distance in the latent space between positive pairs and maximizes the separation from negative samples. This separation is typically accomplished using contrastive loss functions that incorporate arbitrary distance metrics.


In our study, we utilized self-supervised learning techniques to train CXRBase on a vast amount of unlabelled CXR images, leveraging the contrastive principle to learn meaningful representations.

\textbf{Self-prediction}
Self-supervised learning (SSL) involves the process of masking or augmenting parts of an input and using the unaltered portions to reconstruct the original input. The concept of self-prediction SSL originated from the field of Natural Language Processing (NLP), where state-of-the-art models were pre-trained using Masked Language Modeling, predicting missing words in a sentence. This approach achieved significant success in NLP.

In computer vision, early attempts were made to apply similar techniques by masking or augmenting random patches of an image and training Convolutional Neural Networks (CNNs) to reconstruct the missing regions as a pre-training strategy. However, these early attempts had only moderate success. The introduction of Vision Transformers (ViT) \cite{dosovitskiy2021image} allowed computer vision models to adopt the same transformer-based architecture as NLP models. Recent studies, such as BERT Pre-Training of Image Transformers (BEiT) \cite{wang2022image} and Masked Autoencoders (MAE) \cite{he2021masked}, combine ViT with self-prediction pre-training objectives and have achieved state-of-the-art results when fine-tuned on various natural image benchmarks.

In self-prediction SSL, masking or augmentations are applied only to specific portions of the input image, and the remaining unaltered portions are used to guide the reconstruction process. In contrast, generative-based SSL methods apply augmentations to the entire image and subsequently reconstruct the entire image. The main difference lies in the scope of the masking or augmentations applied during the SSL process.

\subsection{Medical Image Analysis}
ConVIRT \cite{zhang2022contrastive} employs a dual-tower architecture that processes image-text pairs and utilizes contrastive loss to extract text-related image features. Building on this foundation, GLoRIA \cite{Huang2021GLoRIAAM} introduces token-level local contrastive loss, enabling finer-grained feature extraction.
MGCA \cite{wang2022multigranularity} enhances the generalizability of learned visual representations through a combination of instance-wise alignment, fine-grained token-wise alignment, and disease-level alignment via contrastive learning. To leverage the rich semantic information inherent in the biomedical domain, BioViL \cite{boecking2022making} constructs domain-specific corpora and language models.
CheXzero \cite{CheXzero2022} advances the classification of pathologies by employing direct inference learning, allowing for classification without reliance on explicit labels. MRM \cite{zhou2023advancing} formalizes the understanding of radiographs and radiology reports as two complementary masked modeling objectives.
Additionally, OmniFM-DR \cite{xu2023learning} utilizes a unified
transformer model specifically designed for multi-modal clinical tasks by incorporating customized
instruction tuning.
Recently, masked autoencoder (MAE)-based models have demonstrated significant scalability and have substantially improved several benchmarks in self-supervised learning \cite{xiao2022delving,xing2023self,bozorgtabar2023amae}.
\section{Code and Data Availability}
Code and pre-trained models are available at \url{https://github.com/MedHK23/MAE_DR}.
The new dataset released in this study can be found at \url{https://github.com/MedHK23/MAE_DR}.

\bibliographystyle{naturemag}
\bibliography{references} 

\begin{figure}[t]
	\centering
        \includegraphics[width=0.95\linewidth]{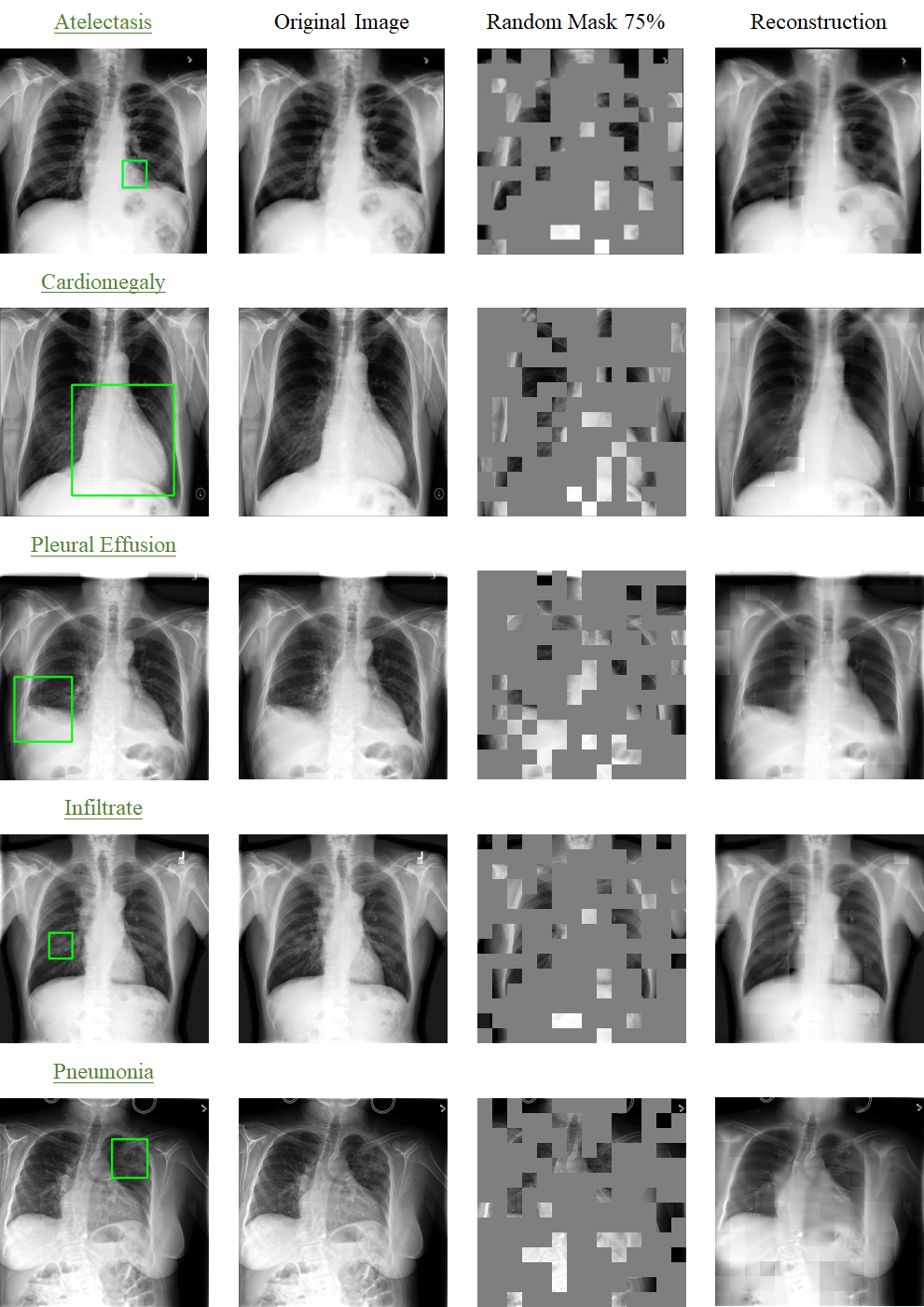}
	\caption{Typical reconstruction examples of \textbf{disease localization task} on several disease categories, i.e., Atelectasis, Cardiomegaly, Pleural Effusion, Infiltrate, Pneumonia.}
	\label{fig:detection}
\end{figure}

\end{document}